\documentclass{ws-procs975x65}
\usepackage{color}

\begin{document}

\title{Study of TeV variability of Mrk 421 from 3 years of monitoring with the Milagro Observatory}

\author{B. PATRICELLI$^*$, M.M. GONZ\'ALEZ AND N. FRAIJA} 

\address{Instituto de Astronom\'ia, UNAM,\\
M\'exico D.F., 04510, M\'exico\\
$^*$E-mail: bpatricelli@astro.unam.mx}

\author{A. MARINELLI}

\address{Instituto de Fisica, UNAM,\\
M\'exico D.F., 04510, M\'exico}

\author{FOR THE MILAGRO COLLABORATION}

\begin{abstract}
The Milagro experiment was a TeV gamma-ray observatory designed to continuously monitor the overhead sky in the 0.1-100 TeV energy range. It operated from 2000 and 2008 and was characterized by a large field of view ($\sim$ 2 sr) and a high duty cycle (\mbox{$\geq$ 90$\%$}). Here we report on the long-term monitoring of the blazar Mrk 421 with Milagro over the period from September 21, 2005 to March 15, 2008. We present a study of the TeV variability of the source and provide upper limits for the measured flux for different time scales, ranging from one week up to one year.
\end{abstract}


\bodymatter

\section{Introduction}\label{sec:intro}
Mrk 421 is one the brightest and closest (z=0.031, see Ref.~\refcite{1991rc3..book.....D}) blazars known. 
It has been a target of many observational campaigns at very high energies (VHE, E $>$100 GeV) with imaging atmospheric Cherenkov telescopes  (IACTs). Its $\gamma$-ray flux has been found to be variable, with flaring episodes over a wide range of timescales, from months down to less than an hour (see, e.g., Ref.~\refcite{2010A&A...524A..48T} and references therein). While IACTs have enough sensitivity to detect short duration flares, extended air shower detectors such as Milagro (see Sec. \ref{sec:milagro}) are better suited to make a continuous monitoring of the flux of the sources, thanks to their high duty cycle. In this paper, we study the TeV variability of Mrk 421 using 3 years (from September 2005 to March 2008) of data from Milagro.

\section{The Milagro Observatory}\label{sec:milagro}
Milagro (see Ref.~\refcite{2004ApJ...608..680A}) was a large water-Cherenkov detector located in the Jemez Mountains near Los Alamos, New Mexico, USA at an altitude of 2630 m above sea level. It operated from 2000 to 2008. It was composed of a central 80 m $\times$ 60 m $\times$ 8 m reservoir instrumented with 723 photomultiplier tubes (PMTs) arranged  in two layers. The top ``air-shower" layer consisted of 450 PMTs under 1.4m of purified water, while the bottom ``muon" layer had 273 PMTs located 6m below the surface. The air-shower layer was used to reconstruct the direction of the air shower by measuring the relative arrival times of the shower particles across the array. The muon layer was used to discriminate between gamma-ray induced and  hadron-induced air showers. In 2004, a sparse 200 m x 200 m array of 175 ``outrigger'' was added around the central reservoir. This array increased the area of the detector and improved the gamma/hadron separation. The experiment was sensitive to extensive air showers resulting from primary gamma rays at energies between 100 GeV and 100 TeV (see Refs.~\refcite{2008ApJ...688.1078A} and  \refcite{2008PhRvL.101v1101A}).

\section{Variability}
\begin{figure}
\begin{center}
\epsfig{file=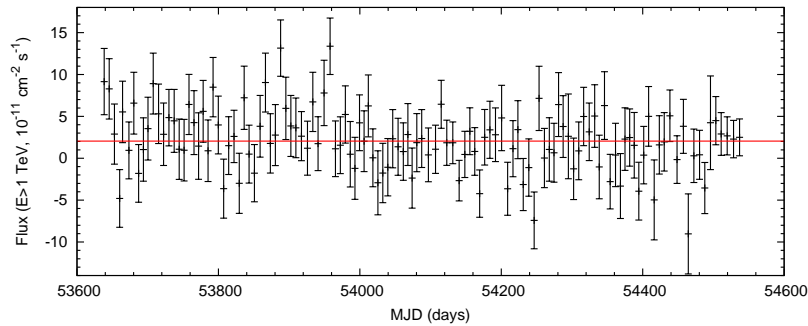,width=4.2in}
\caption{Light curve of Mrk 421 (black points); the red solid line represents the average value of the flux above 1 TeV: $\bar f$=($2.052 \,\pm 0.304$) $10^{-11}\,\rm{cm^{-2}\,s^{-1}}$. Each bin represents $\sim$ 1 week of data.}
\label{fig:LC}
\end{center}
\end{figure}

We analysed Milagro data collected from September 21, 2005 to March 15, 2008, period over which Mrk 421 was detected with a statistical significance of 7.1 standard deviations, at a median energy of 1.65 TeV. In Fig.~\ref{fig:LC} it is shown the Milagro light curve (LC) above 1 TeV for Mrk 421. We chose a time binning of $\sim$ 1 week \footnote{Milagro data are recorded in tapes and each tape contains data collected over a time interval that, on average, is of $\sim$ 1 week; each time bin in the light curve corresponds to data recorded in one tape.}  
in order to have high statistics, as well as to minimize the effects of the detector instabilities. It can be seen that the flux is consistent with being constant along the 3-year monitoring period, with an average value of $\bar f$= ($2.052 \,\pm 0.304$) $\times 10^{-11}\,\rm{cm^{-2}\,s^{-1}}$ ($\chi^2$=134 for 122 degrees of freedom, which gives a $\chi^2$ probability of 21.1\%). Therefore, from the light curve it is clear that there is no evidence for flares in the data. To confirm this, we also calculated the significance above the average flux for each time bin, finding that all the Milagro measurements have a significance less than 3 standard deviations (after the correction for the number of trials). 

\begin{figure}
\begin{center}
\epsfig{file=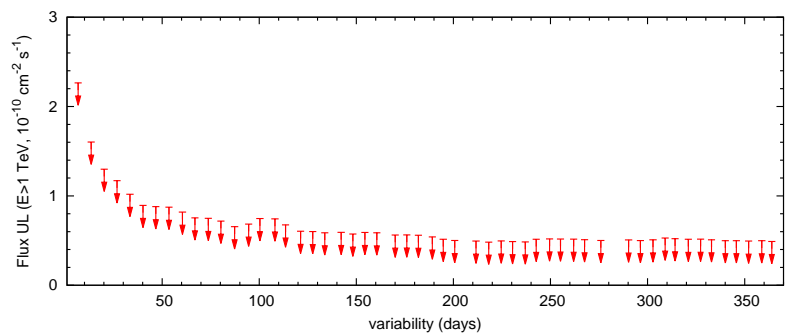,width=4.2in}
\caption{Upper limits on the flux as a function of flare duration.}
\label{fig:UL}
\end{center}
\end{figure}

As we found no evidences of flares, we computed the upper limit $F_{\rm UL}$ on the measured flux considering different time scales, from one week to $\sim$ 12 months, as for Mrk 421 outbursts lasting up to several months have been observed (see e.g. Ref.~\refcite{2010A&A...524A..48T}). Then we chose a confidence level of 99.7 $\%$ and used the method of Helene\cite{1983NIMPR.212..319H}. The results are shown in Fig.~\ref{fig:UL}. It can be seen that $F_{\rm UL}$ varies from \mbox{2.26 $\times$ 10$^{-10}$ cm$^{-2}$ s$^{-1}$} to 0.48 $\times$  10$^{-10}$ cm$^{-2}$ s$^{-1}$; its maximum value is found for a time scale of $\sim$ one week. 

The long-term average flux here reported, together with the VHE flux state distribution of Mrk 421 (see e.g. Ref.~\refcite{2010A&A...524A..48T}), is fundamental to calculate the TeV duty cycle of the source and will be object of a future work. A further extension of the work will be also the calculation of the upper limits on the flux as a function of time to make a comparison with observations from other VHE experiments in the same period as Milagro.

\section*{Acknowledgments}
We gratefully acknowledge Scott Delay and Michael Schneider for their dedicated efforts in the construction and maintenance of the Milagro experiment. This work has been supported by 
the Consejo Nacional de Ciencia y Tecnolog\'ia (under grant Conacyt 105033), Universidad Nacional Aut\'onoma de M\'exico (under grant PAPIIT IN105211) and DGAPA-UNAM.

\bibliographystyle{ws-procs975x65}
\bibliography{bib}

\begin{thebibliography}{1}

\bibitem{1991rc3..book.....D}
G.~{de Vaucouleurs et al.}, {\em {Third Reference Catalogue of Bright
  Galaxies.}} (Springer, New York, USA, 1991).

\bibitem{2010A&A...524A..48T}
M.~{Tluczykont et al.}, {\em A\&A} {\bf 524}, p. A48 (2010).

\bibitem{2004ApJ...608..680A}
R.~{Atkins et al.}, {\em ApJ} {\bf 608}, 680 (2004).

\bibitem{2008ApJ...688.1078A}
A.~A. {Abdo et al.}, {\em ApJ} {\bf 688}, 1078 (2008).

\bibitem{2008PhRvL.101v1101A}
A.~A. {Abdo et al.}, {\em Physical Review Letters} {\bf 101}, p. 221101 (2008).

\bibitem{1983NIMPR.212..319H}
O.~{Helene}, {\em Nuclear Instruments and Methods in Physics Research} {\bf
  212}, 319 (1983).

\end{thebibliography}

\end{document}